\documentstyle[12pt]{article}
\evensidemargin \oddsidemargin
\catcode `\@=11
\def\numberbysection{\@addtoreset{equation}{section}
         \renewcommand{\theequation}{\thesection.\arabic{equation}}}
\def\be{\begin{equation}}
\def\ee{\end{equation}}
\newcommand{\ba}{\begin{eqnarray}}
\newcommand{\ea}{\end {eqnarray}}
\newcommand{\nn}{\nonumber}
\def\ri{\right}
\def\le{\left}

\def\a{\alpha}
\def\b{\beta}
\def\g{\gamma}
\def\G{\Gamma}
\def\d{\delta}
\def\D{\Delta}

\def\sp{\sigma^+}

\def\sx{\sigma^x}
\def\sy{\sigma^y}
\def\sz{\sigma^z}

\def\w{\omega}
\def\nil{\emptyset}

\def\lra{\longrightarrow}

\def\pa{\partial}
\def\half{\frac{1}{2}}

\numberbysection
\begin{document}
%
\pagestyle{empty}
\vspace* {10mm}
\renewcommand{\thefootnote}{\fnsymbol{footnote}}
\begin{center}
\large
   {\bf FINITE-SIZE SCALING STUDIES}\\[4mm]
   {\bf OF REACTION-DIFFUSION SYSTEMS}\\[4mm]
   {\bf Part III: Numerical Methods}\\[1cm]
\end{center}
\begin{center}
\normalsize
   Klaus Krebs, Markus Pfannm\"uller, Horatiu Simon and Birgit Wehefritz
\\[5mm]
   {\it Universit\"{a}t Bonn,
   Physikalisches Institut \\ Nu\ss allee 12,
   D-53115 Bonn, Germany}
\\[2cm]
{\bf Abstract}
\end{center}
\renewcommand{\thefootnote}{\arabic{footnote}}
\addtocounter{footnote}{-1}
%
%
\small

The scaling exponent and scaling function for the 1D single species
coagulation model $(A+A\rightarrow A)$ are shown to be universal, i.e.
they are not influenced by the value of the coagulation rate. They are
independent of the initial conditions as well. Two different numerical
methods are used to compute the scaling properties:
Monte Carlo simulations and extrapolations of exact finite lattice data.
These methods are tested in a case where analytical results are available.
It is shown that Monte Carlo simulations can be used to compute even the
correction terms. To obtain reliable results from finite-size extrapolations
exact numerical data for lattices up to ten sites are sufficient.
\\
\rule{5cm}{0.2mm}
\begin{flushleft}
\parbox[t]{3.5cm}{\bf Key words:}
\parbox[t]{12.5cm}{Reaction-diffusion systems, finite-size scaling,
                 Monte Carlo simulations,
		 non-equilibrium statistical mechanics,
                 coagulation model}
\\[2mm]
\parbox[t]{3.5cm}{\bf PACS numbers:}
\parbox[t]{12.5cm}{05.40.+j, 05.70.Ln, 82.20.Mj, 82.20.Wt}
\end{flushleft}
\normalsize
\vspace{3cm}
\begin{flushleft}
BONN HE-94-02\\
cond-mat/9402019\\
Bonn University\\
January 1994\\
ISSN-0172-8733
\end{flushleft}
\thispagestyle{empty}
\mbox{}
\newpage
\setcounter{page}{1}
\pagestyle{plain}
%
%
%
\section{Introduction}
\hspace{\parindent}This article  completes our investigations of the
finite-size
scaling properties of reaction-diffusion systems. Before getting into details
we would like to outline a couple of questions which we will discuss in the
present work:
\begin{itemize}
\item \underline{Universality of the scaling properties}\\
To what extend are
the scaling exponents and functions and the corrections influenced by
special values of the reaction rates?
\item \underline{Influence of the initial conditions}\\
Does the choice of the initial
condition change the scaling exponents or functions or does it enter only in
the corrections?
\item \underline{Numerical methods}\\
Is it practically possible to compute the scaling properties
by Monte Carlo simulations? Can extrapolations of exact finite lattice data
provide information about the scaling exponents and functions or even about
the corrections? Exact numerical diagonalisations of the
quantum chain Hamiltonians corresponding to a certain model
are only possible for very small lattices. Typically one can handle
about 20 sites for a two state and 10 sites for a three state model.
Therefore a point of major interest is to figure out what sizes
of lattices are necessary to obtain reliable results. Furthermore we discuss
which parameters such as boundary conditions, initial conditions,
reaction rates or paricular values of the scaling variable $z$
are most suitable for numerical investigations.
\end{itemize}

We choose the single species coagulation model as an example to study these
questions. In the previous two papers (hereafter referred
to as I and II) we used analytical methods to investigate this model with
periodic \cite{Pap1} and with open \cite{Pap2} boundary conditions and for a
very special value of the coagulation rate. We have
shown that in the finite-size scaling limit
$L \rightarrow\infty$, $t \rightarrow \infty$
with the scaling variable $z=\frac{4Dt}{L^2}$ kept constant the
concentration of particles takes the form:
\be
c(z,L)=L^x[F_0(z)+L^{-y}F(z)+\cdots]\;.
\label{eq:scf}
\ee
The scaling exponent $x$ was found to be equal to $-1$ for both kinds of
boundary conditions. The scaling functions $F_0(z)$ however are found
to be different for periodic and for open boundary conditions.

We briefly review the definition of the coagulation model (for details we refer
to paper I).
The model is defined on an one-dimensional lattice of length $L$. The following
reactions are allowed:
\begin{itemize}
\item diffusion with rate $D$
\be
A+\nil \longleftrightarrow \nil+A\;,
\ee
\item coagulation with rate $c$
\be
A + A \longrightarrow A\;.
\ee
\end{itemize}
We always choose units of time so that $D=1$. In \cite{Alcaraz} it was
shown that the master equation describing a recation-diffusion system can
be mapped onto an euclidean Schr\"odinger equation. Then the time evolution
of the system is determined by the Hamiltonian of an one-dimensional
quantum chain:
\be
\frac{\pa}{\pa t}|P\rangle = -\tilde{H}|P\rangle\;.
\ee
In paper I we have shown that the coagulation model is described by the
Hamiltonian $\tilde{H}=H_0+H_1$ where:
\be
H_0  =  -\half\sum_{i=1}^{L}
           \Bigl[\sx_i\sx_{i+1}+\sy_i\sy_{i+1}+\D'\sz_i\sz_{i+1}
          +(1-\D')(\sz_i+\sz_{i+1})+\D'-2\Bigr]
\ee
and
\be
H_1  =  -\half(1-\D')\sum_{i=1}^{L}
          [\sp_i(1-\sz_{i+1})+(1-\sz_i)\sp_{i+1}] \;.
\label{eq:HC1}
\ee
By:
\be
\D'=1-c
\ee
we denote the asymmetry between the diffusion and the coagulation rate.
This notation has already been used in paper~I and \cite{Alcaraz}.
The spectrum is determined by $H_0$ alone, which is the Hamiltonian for the
XXZ-chain in an external magnetic field. As long as no decoagulation
reaction $(A\rightarrow A+A)$  occurs
the system is on the Pokrovsky-Talapov line \cite{Alcaraz} independently of
$\D'$.

The first aim of the present paper is to investigate the scaling behaviour
(\ref{eq:scf}) of the concentration for
different choices of the coagulation rate $c$, i.e. different values of $\D'$.
It can be shown
\cite{Henkel_unp} that in the continuum, along the Pokrovsky-Talapov line,
the system is always massless with the same quadratic dispersion relation.
We expect that this fact is reflected in the scaling properties.
For $\D'=0$ the spectrum can be obtained in terms of free fermions. This
fact allows the analytical solutions presented in paper~I and II.
For $\D'\not=0$ these calculation do not apply. Therefore we use
Monte Carlo simulations to determine the scaling exponent and function as
well as the corrections.

Our second point of interest is to figure out in which way the choice of
initial
conditions influences the scaling properties. For $\D'=0$ the case of
uncorrelated initial conditions with arbitrary initial occupation probability
$p$ was treated analytically in paper I and II. Now we use
Monte Carlo simulations and extrapolations from finite lattices to study
initial states with small clusters. These studies are carried out for
$\D'=0$ only.

We test the accuracy of Monte Carlo simulations in the case of
$\D'=0$ where analytical results are available.
The comparison of Monte Carlo data and analytic results demonstrates that
Monte Carlo simulations provide reliable information about the scaling
exponents and functions as well as about the corrections.

It has been pointed out in paper I that scaling relations allow to compute
estimates for
critical exponents of an infinite system from the scaling behaviour of a
finite one. An important purpose of the present paper is therefore to
figure out what sizes of lattices are necessary to compute the scaling
properties numerically.
The case of $\D'=0$ is used
to test the accuracy of finite lattice extrapolations.
It is shown that finite-size analysis allows to determine the scaling
properties already from numerical data for lattices of length smaller than
ten sites.

We decided to organize the paper according to the different numerical
methods we use. In Section \ref{sec:MC}
we discuss Monte Carlo simulations of reaction-diffusion
systems and present the results obtained by this method. Section
\ref{sec:bst} is devoted to extrapolations from finite
lattices. The accuracy of the extrapolations is studied and the
investigation of different initial conditions in the case of open
boundary conditions is presented. We close with the discussion of our results.
%
%
\section{Monte Carlo Simulations}
\label{sec:MC}
\hspace{\parindent}Before we present the results of the Monte Carlo
simulations we briefly explain how the simulations are performed.
The dynamics of the system are determined by the
rates $\G_{\a,\b}(\g,\d)$
for the nearest neighbour interactions:
\be
(\a,\b) \lra (\g,\d) \;.
\ee
Here a value of $\a,\b,\g,\d=1$ corresponds to a particle and a value of 0
to a vacancy. (This notation is connected to the one used in paper I by
$w_{\a-\g,\b-\d}(\g,\d)=\G_{\a,\b}(\g,\d)$, the change is only made for
convenience.)

We consider two types of initial conditions: in an uncorrelated initial
state all sites are occupied with equal probability $p$ at $t=0$;
a weakly correlated state consists of an interpolating sequence of
two particles and two vacancies ($AA\nil\nil AA\nil\nil\cdots$).

We only study
vacuum driven processes, i.e. no reactions occur on a pair of empty sites.
Therefore we use the ``direct method'' \cite{Argyrakis,ben-Avraham} for our
simulations.
By $\D t$ we denote a discrete time step. Each Monte Carlo step consists
of the following operations:
\begin{itemize}
\item[1.)]
A particle is chosen at random and afterwards, with equal probability, one
of its neighbours. The resulting pair $(\a,\b)$ cannot consist of two
vacancies.
\item[2.)]
A  new configuration $(\g,\d)\not=(\a,\b)$ is chosen with the probability:
\be
n(\a,\b) \G_{\a,\b}(\g,\d)\D t
\ee
or the pair remains unchanged with probability:
\be
1-n(\a,\b)\D t\sum_{(\g,\d)\not=(\a,\b)}\G_{\a,\b}(\g,\d)\;.
\ee
Here $n(\a,\b)$ is defined by:
\be
n(\a,\b)=\le\{\begin{array}{ccc}
 1 & \mbox{if} & (\a,\b)=(1,1)\\
 2 & \mbox{if} & (\a,\b)=(1,0)\\
 2 & \mbox{if} & (\a,\b)=(0,1)\end{array}\ri.\;.
\ee
This factor \cite{Doering} simply accounts for the fact that a pair containing
one vacancy is chosen with half the probability of choosing one that consists
of two particles.
\item[3.)]
Finally the time is increased:
\be
t \lra t+\D t/N
\label{eq:time}
\ee
where $N$ is the total number of particles on the lattice.
\end{itemize}
If the coagulation process has been chosen in a Monte Carlo step, the average
concentration decreases by an amount of:
\be
q_c=\frac{1}{L N_s}
\ee
where $N_s$ denotes the number of samples (i.e. program runs) used in the
simulation. Therefore the simulated time $t$ should not reach values
where the concentration $c(t)$ becomes of the order of $q_c$.
Because we are interested in the scaling behaviour and the corrections, it
is necessary to perform the simulations on small lattices ($L<100$). Therefore
the number of samples $N_s$ has to be large. A large number of samples
is advantegous because the standard deviation of the mean decreases like
$(N_s)^{-0.5}$. The limit is set by the growth of the necessary CPU time.
Averaging over $N_s=20,000$ samples provides very accurate results as the
test described in the sequel will show.

The time discretisation $\D t$ clearly influences the quality of the
simulations. For smaller lattices this influence grows,
because the number of particles is small (cf. Eq. (\ref{eq:time})).
As a test, we perform simulations for $\D'=0$ and periodic boundary
conditions and compare the results with the exact expressions from
paper I. The difference of the concentration averaged over 20,000
program runs and the  exact value is used as a measure for the accuracy
of the simulation. Tab. \ref{tab:taex} shows the maximum value of the
relative deviation which is observed during simulations up to a value of
$z_{\mbox\scriptsize{max}}=0.1$. Then the maximum simulated time is
$t_{\mbox\scriptsize{max}}=z_{\mbox\scriptsize{max}}L^2/4$.
It can be seen that $\D t$ has to be decreased to obtain results with equal
accuracy for smaller lattices.
The table shows typical values for $\D t$ which are used for
different lattice lengths.

However, for very small values of $\D t$ the probability that ``something''
will happen in a Monte Carlo step becomes very small in comparison with
the probability that ``nothing'' will happen. Therefore the fact that a random
generator never provides perfectly uniformly distributed numbers limits
the possibility to improve the simulations by decreasing $\D t$.

Another serious limitation is the growth of the necessary CPU time which
is proportional to $(\D t)^{-1}$. As an example take the CPU time of
80 seconds for $L=3$, $N_s=20,000$, $\D t=5\cdot10^{-3}$ and
$z_{\mbox\scriptsize{max}}=0.1$.

We are interested not only in the scaling function but in the corrections
as well. The latter can only be determined for values of the scaling
variable $z$
for which the corrections are much larger than the numerical errors. Values
of $z=0.01,\ldots,0.1$ turn out to be suitable.

To demonstrate the accuracy we make simulations for
$L=9,10,\ldots,16$ with $\D t=5\cdot10^{-3}$ and averaged over 20,000 samples.
The difference between the
simulated concentration and the
exact expression is always less than 0.4\% and less than the standard
deviation of the mean. With the help of a $\chi^2$-fit we compute
approximations for the scaling function and the corrections in $\frac{1}{L^2}$
for fixed values of $z$. The results are given in Tab. \ref{tab:test}  along
with the exact expressions from paper I. It can clearly be seen that
Monte Carlo simulations provide reliable information about the scaling
properties.

The programs were written in Fortran using the RAN random number generator and
executed on DEC work-stations.

\subsection{The Influence of the Coagulation Rate}
\hspace{\parindent}To solve the coagulation model exactly in paper I and II
it was necessary to choose the coagulation rate equal to the diffusion rate,
i.e. $\D'=0$. Now we use Monte Carlo simulations to study the way the
scaling properties change for different choices of the coagulation rate,
i.e. $\D'\not=0$.

We use a full lattice as initial configuration corresponding to an
occupation probability of $p=1$. Both, periodic and open boundary conditions
are considered. Averages are always taken over $N_s=20,000$ samples,
except where other numbers are stated.

The first step is to determine the scaling exponent $x$ in Eq. (\ref{eq:scf}).
For this purpose, for each value of $\D'$,
we perform simulations for $L=500,550\ldots950$  taking
$\D t=0.5$. Averaging over $400$ samples is sufficient in this case,
whereas in all other simulations we use $20,000$ samples. For each value of
$z$, $x$ is determined with the help of the $\chi^{2}$ test,
according to:
\be
\log\:(c(z))=\mbox{const.}+x\:\log(L) \;.
\ee
With an accuracy of 5\% or better we obtain $x=-1$, independently of $\D'$.

The second step is to find the leading term of
the corrections. In order to do so, we need an approximation for the
scaling function $F_0$. For this purpose we use the value of $L c(z,\D')$
obtained from simulations on a lattice of length $L=1000$.
Then we perform another set of simulations on lattices of length
$L=30,35\ldots85$ with $\D t=5\cdot 10^{-3}$.
Using this second set of data, we make a logarithmic fit of the type
\be
\log\:(L\:c(z)-F_0)=\mbox{const.}-y\:\log(L) \;.
\label{eq:eq5.11}
\ee
In all cases the correction exponent $y$ is found to be equal to $+1$ with
an error of 15\% or less. Observe the contrast to
the case of $\D'=0$ where the leading correction exponent is
$y=2$. {\em So the dominant correction for $\D'\not=0$ is
of the order of $1/L$.}

Therefore we finally approximate the second set of data by:
\be
L\:c(z,\D')=F_0(z,\D')+\frac{F_1(z,\D')}{L}+\frac{F_2(z,\D')}{L^2}\;.
\label{eq:f}
\ee

{\em The scaling function $F_0$ is found to be independent of $\D'$ }
and has the values given in paper I, respectively II, i.e.
$F_0(z,\D')=F_0(z,0)$. This can be seen in the Tabs. \ref{tab:ts} and
\ref{tab:fts}, where the values obtained for $F_0$ are given for periodic
and for open boundary conditions.

The first order correction function $F_1$  behaves according to the following
scheme:
\ba
F_1(z,\D') & > &0\:\:\mbox{for}\;\; \D'>0\nn\\
F_1(z,\D') & = &0\:\:\mbox{for}\;\; \D'=0\nn\\
F_1(z,\D') & < &0\:\:\mbox{for}\;\; \D'<0\;.
\label{eq:scheme}
\ea

The values for the first and second correction functions are given in Tabs.
\ref{tab:tfc} and \ref{tab:tsc} for periodic boundaries and in Tabs.
\ref{tab:ftfc} and
\ref{tab:ftsc} for open boundaries, together with the analytic
results for $\D'=0$. As the second
order correction term, i.e. $F_2(z,\D')/L^2$, becomes already of the order of
the statistical errors of the simulations the correction function $F_2$ can
only be determined with large numerical errors.
The correction functions are presented
in Figures 1-4.

As has been explained in paper I the scaling limit for small $z$ and
the long-time behaviour in the thermodynamic limit ($L \rightarrow\infty$) are
related. For periodic boundaries and $\D'=0$ we found the following expansion
for the concentration:
\be
c(t) = \sqrt{\frac{1}{2\pi t}}
       \le[1-\frac{1}{16 t}\le(\frac{(p-2)^2}{p^2}-\half\ri)+\cdots\ri]\;.
\ee
The leading term is obtained from the scaling function and the second term
from the expansion of the corrections in $1/L^2$. An analysis of the
leading corrections for $\D'\not=0$ which are of the order of $1/L$ and
behave like $1/z$ leads to an expansion:
\be
c(t) = \sqrt{\frac{1}{2\pi t}}\le[1+\frac{a(\D',p)}{\sqrt{t}}
                             +\cdots\ri]\;.
\ee
Thus the change of the coagulation rate does not affect the leading term
of the large time expansion of the concentration in the thermodynamic limit.
However the next-to-leading term is no longer of the order of
$t^{-\frac{3}{2}}$ as for $\D'=0$ but of the order of $t^{-1}$.
The coefficient $a$ depends on the value of $\D'$ and on the initial
condition. For $p=1$ and periodic boundary conditions we present the
values of $a$ for different $\D'$ in Tab. \ref{tab:ltime}. These data
are computed using a $\chi^2$-fit of the values for the first correction
function $F_1$ (cf. Tab. \ref{tab:tfc}).

If one is interested only
in the time dependence for systems in the thermodynamical limit
$(L\rightarrow\infty)$ there is of course no need to to calculate
the scaling expansion first. Instead one can directly perform
simulations on large lattices. For $\D'=0.5$ we check the result
given in Tab. \ref{tab:ltime} by a simulation on a lattice of
2000 sites. A $\chi^2$-fit of the data averaged over $N_s=20.000$
samples gives $a(0.5, 1)=0.818 \pm 0.006$. This value is compatible
with the one obtained from the scaling data.
However the errors for the direct simulations are smaller by one order
of magnitude.
\subsection{The Influence of the Initial Conditions }

\hspace{\parindent}For periodic boundaries and  $\D'=0$ we use Monte Carlo
simulations to study the influence of a weakly correlated initial state.
The configuration at $t=0$ is chosen to
be a interpolating sequence of two particles and two vacancies
($AA\nil \nil AA \nil \nil \ldots$).

Lattice lengths of $L=36,40\ldots 80$ ($\D t=5\cdot 10^{-3}$) are used.
We analyse the data using the procedure described in the previous section.
The results, shown in Tab. \ref{tab:icc}, confirm
{\em that the scaling function is not influenced by an initial state with
small clusters}. The corrections  are still
of the order $\frac{1}{L^{2}}$,  but they are different from those given in
paper I for $p=0.5$, although the initial concentration is the same,
$c(0)=0.5$.

%
\section{Extrapolation from Finite Lattices}
\label{sec:bst}
\hspace{\parindent}The coagulation-decoagulation model, being exactly
solvable,  provides a good experimental field to test the accuracy of
numerical extrapolation from finite lattices to determine scaling functions and
exponents.
Our special interest is to find out from what size of lattices it is possible
to obtain reliable information about the scaling exponent and the scaling
function of the particle concentration as well as about the corrections.

A common method used to perform a finite-size extrapolation was proposed
by Bulirsch and Stoer \cite{Bulirsch} (in the sequel being referenced
as BST).

We first shortly review this algorithm, following \cite{Malte}.
Let $T(\frac{1}{L})$ be a function with an expansion
\be
T\left(\frac{1}{L}\right)\:=\: T+\:\left(\frac{1}{L}\right)^{\w}\:T' \:+\:
\left(\frac{1}{L}\right)^{2\w}
\:T''\:+\ldots
\ee
where we are interested in the asymptotic behaviour in the limit $L\rightarrow
\infty$. A table of extrapolants is formed
\be
\begin{array}{ccc}
T_0^{(0)} & & \\
 & T_1^{(0)} & \\
T_0^{(1)} & & T_2^{(0)}\\
 & T_1^{(1)} & \\
T_0^{(2)} & &
\end{array}
\ee
by application of the following rules
\ba
T_{-1}^{(L)} & = & 0\nn\\
T_0^{(L)} & = & T\left(\frac{1}{L}\right)\nn\\
T_m^{(L)} & = & T_{m-1}^{(L+1)}\:+(T_{m-1}^{(L+1)}-T_{m-1}^{(L)})
\le[\le(\frac{L+m}{L}\ri)^{\w}\le(1-\frac{T_{m-1}^{(L+1)}-T_{m-1}^{(L)}}
                           {T_{m-1}^{(L+1)}-T_{m-2}^{(L+1)}}\ri)-1\ri]^{-1}\:.
\ea
Here $\w$ is a free parameter. The algorithm arises by approximating
the function $T\left(\frac{1}{L}\right)$ by a sequence of rational functions
in the variable
$(\frac{1}{L})^{\w}$. $T^{(L)}_1$ is expected to converge faster than
$T^{(L)}_0$ to the limit $T(0)$ when $L\rightarrow\infty$.
The parameter $\w$ is determined by the condition that the errors of the
approximation, i.e. the difference of two successive extrapolants
$T^{(L+1)}_m-T^{(L)}_m$ are minimized for each $m$.
This method is very convenient for our purpose, because a high precision for
the extrapolated functions is obtained using only a few input values (which,
however, must be known very precisely).

\subsection{Extrapolations for Periodic Boundary Conditions}
\hspace{\parindent}As we are interested to know how good the
extrapolation from small lattices are, we
make calculations for the scaling function, the scaling exponent, the
correction
exponent and the correction function for the coagulation model with periodic
boundary conditions and for $\D'=0$.
For the extrapolations, we take lattice lengths between 2 and 9 sites.
We tried to work with seven input values ($L=2,3,4,5,6,7,8$)
and with eight input values ($L=2,3,4,5,6,7,8,9$). For small values
of $z$ the algorithm is convergent only in the latter case.
The input values are calculated using the exact expression for the
concentration in the coagulation model derived in paper I (cf. eq. (7.2)).

For $z$ the values $z=0.1,0.2,\ldots,1.0$ were chosen, covering a sufficiently
large range in
time between $t=0.1$ and $t=25$ where the behaviour of the scaling
function and its corrections are investigated.

Let us now describe the extrapolations in detail:
\begin{itemize}
\item[a)] \underline{Scaling exponent}

In order to get a higher precision in our estimates, we write the scaling
exponent as $x=-1+\tilde{x}$ and try to determine
the difference $\tilde{x}$ to the theoretical value for the scaling exponent
$x=1$, using the following expression to form the extrapolants:
\be
\tilde{x}_{0}^{(L)}\:=\:
\frac{\ln(\frac{(L+1)\;c(L+1,z)}{L\;c(L,z)})}{\ln(\frac{L+1}{L})}\:.
\ee
The only inconvenience of this method is the fact that one obtains one
input value less than for the extrapolation of the scaling function.
The results are shown in Tab. \ref{tab:extra} for each $z$. We find
$\tilde{x}=0$ with a numerical error
less than $3\cdot10^{-3}$.

\item[b)] \underline{Scaling function}

The scaling function $F_{0}-1$ is extrapolated from
\be
T_{0}^{(L)} \,=\,L\:c(z,L)-1\:.
\ee
The results
are compared to the exact values of the
scaling function $F_{0}(z)-1$ calculated from paper I in Tab.
\ref{tab:extra}. The bigger the values of $z$ are,
the higher the precision of the
extrapolants becomes: for small $z$, the difference to the theoretical
value is of
the order of $2\cdot 10^{-3}$, for $z=1.0$ we even get a precision of 10
digits in
comparison to the exact scaling function. This can easily be understood
looking at
the definition of $z=\frac{4Dt}{L^2}$. A higher value of $z$ means that the
time $t$ is already much later and that we are closer to the finite-size
scaling
regime where $t\rightarrow\infty$. So the values of $L\;c(L,z)$ for bigger $z$
should better approximate the theoretical scaling function.

\item[c)]\underline{Correction exponent}

Similar to a), we extrapolate the difference
$\tilde{y}=2+y$ to the theoretical correction exponent $y=-2$. The input values
are calculated according to
\be
\tilde{y}_{0}^{(L)}\:=\:\frac{\ln(\frac{(L+1)^{2}[(L+1)c(L+1,z)-F_{0}(z)]}
{L^{2}[L\;c(L,z)-F_{0}(z)]})} {\ln(\frac{L+1}{L})}\:.
\ee
For the value of the scaling function $F_{0}(z)$ we use the result obtained in
b). So there is already a numerical error in the input values leading to
a lower precision for the correction exponent in comparison to the scaling
exponent.
We find $\tilde{y}=0$ with an error of $10^{-3}$. The extrapolated value at
$z=0.6$ is worse because the correction function $F_2(z)$ is close to its
zero.

\item[d)]\underline{Correction function}

We are interested in the corrections $F_{2}(z)$ to the scalingfunction, i.e.
we extrapolate
\be
T_{0}^{(L)}\:=\: L^2\:\left(L\;c(L,z)-F_{0}(z)\right)\:.
\ee
For $F_{0}(z)$ we again use the result obtained in b).
Arguing similarly to the approximation of the scaling function, we can
understand the high precision of the extrapolated correction function for big
values of $z$. But even for small values of $z$ (except for $z=0.1$)
the precision
is in the range of $6\cdot10^{-4}$. The bad value for the extrapolated
corrections for $z=0.1$ can be understood looking at the corresponding
extrapolated scaling function which is known only with a precision of
$2\cdot10^{-3}$. As this value enters the input values for the extrapolation of
the corrections, the algorithm cannot lead to a reasonable result.
\end{itemize}

In conclusion, we can say that the BST algorithm provides a powerful tool
for extrapolations from finite lattices. Already from ten sites, for which
the Hamiltonian can easily be diagonalized explicitly in order to get
a value for the concentration, we obtain extrapolations with a very high
precision. This knowledge can be used especially to examine non-integrable
chemical models and to determine their critical exponents and finite-size
scaling properties.

\subsection{Extrapolations for Open Boundary Conditions}
\hspace{\parindent} Once having tested the accuracy of the BST-algorithm
we now use it to examine the influence of a weakly correlated
initial state
on the scaling and the correction function for open boundary conditions.
For this purpose, we numerically calculate the concentration
for finite lattices
using the expressions found in paper II for the eigenfunctions
of the system of differential equations describing the time evolution of the
coagulation model (cf. eqs. (2.13)-(2.16)).
Analytically, these calculations become too cumbersome.
Afterwards, we extrapolate the finite lattice data with BST. This time we
use for the input values lattice lengths between $20$ and $40$ sites in order
to get a higher precision for the extrapolated values.

Using the same procedure as described in the previous section, we determine
the scaling function and the corrections for the weakly
correlated initial condition
$AA\nil \nil AA \nil \nil \ldots$.
The results are shown in Tab. \ref{tab:openex} in comparison
to the analytical values for uncorrelated initial conditions (initial
occupation probability $p=0.5$).

The scaling function is again found to be independent of the initial
conditions with a precision of $6$
digits for values of $z$ larger than $z=0.1$.

On the other hand, the corrections are influenced by the different
choice of initial
conditions. They are even different for uncorrelated and weakly correlated
initial conditions with the same initial concentration. The corrections
are smaller for the interpolating sequence of particles and holes than for
the random distribution in the case of open boundary conditions.
%
%
%
\section{Conclusions}
\hspace{\parindent}
In this paper, we presented the numerical examination of the finite-size
scaling behaviour of the coagulation model.

We have shown on the one hand the universality of the scaling function
with the help of Monte Carlo simulations.
On the other hand, we were able to determine critical exponents already
from lattices of 10 sites using finite-lattice extrapolations.
The test of these two numerical methods (Monte-Carlo simulations and a
finite-lattice extrapolations) revealed that they can be
successfully applied to characterize the finite-size scaling behaviour of
reaction-diffusion processes.

Let us discuss our results in detail.
\begin{itemize}
\item \underline {Universality of the scaling function and the corrections}

As a first application of the Monte Carlo simulations, we examined
the scaling function for periodic
and open boundary conditions for different tuning of the rates in order to
answer the question whether the scaling function is universal (i.e. independent
of the details of the model): we varied the parameter
$\D'$ that reflects the difference between the diffusion
rate ($D$) and the coagulation rate ($c$). The scaling function was found
to be universal.

As far as the corrections to the scaling function are concerned, we obtained
a leading correction term of the order
$\frac{1}{L}$ in the case of $\D'\,\not=\,0$.
The corresponding correction function  (the coefficient of
$\frac{1}{L}$) is different for different values of
$\D'$. However, it has always the same sign as $\D'$ (cf. Eq. (\ref
{eq:scheme})). Especially, for $\D'\,=\,0$ it is zero as well.
This explains the fact that we obtained leading corrections of the order
$\frac{1}{L^2}$ in the case $\D'\,=\,0$ which was treated analytically
in paper I and II.

\item \underline {Influence of the initial conditions}

We examined the influence of weakly correlated initial states both
for open and for periodic boundary conditions. A configuration with
small clusters $(AA\nil\nil AA\ldots)$ was used as initial state.
We have shown that the
scaling function is not influenced by the initial conditions.
So we observe again the phenomenon of self-organisation. The system
develops according to its intrinsic dynamics and is independent of the
initial conditions for large times and large lattices.

\item \underline {Test of numerical methods}

In order to test the accuracy of Monte-Carlo simulations, we compared
the results obtained by simulation to the data that are analytically
known. This comparison revealed that the Monte-Carlo method provides
reliable results that approximate very well the exact expressions.
So this method can successfully be used to investigate the finite-size
scaling behaviour of chemical models.

As far as finite lattice extrapolations are concerned, we have demonstrated
that the algorithm of Bulirsch and Stoer \cite{Bulirsch} is a useful method to
determine
the scaling and the correction exponents and -functions already from very
small lattices. The only condition
one has to impose is that the input values (for small lattices) have to be
known with a sufficiently high precision.
The accuracy of the extrapolated values depends on the number of input values.
However, in general it is sufficient to take the values for $8-10$ sites
as input values to determine the exponents with a precision of at least
$10^{-3}$.
So this method is very suitable for the analysis of the finite-size
scaling regime for models that cannot be treated analytically.
In this case a numerical diagonalisation is still possible for small
lattices and can be used to generate the input values for the
BST-extrapolation.

\item \underline {Best choice of the parameters for the numerical examination}

The analysis of the coagulation model allows to decide which is the best
choice of the parameters (tuning of the rates, initial conditions, boundary
conditions and value of the scaling variable $z$) for the numerical
determination of the finite-size scaling data.

\begin{enumerate}

\item {\em Tuning of the rates}

As far as this model is concerned, the scaling function can best be determined
if the rates are tuned in a way that $\D'\:=\:0$. Then the
leading correction term is of the order $L^{-2}$, i.e. small for large
$L$; furthermore the correction function (the coefficient of $L^{-2}$)
is smallest for $\D'\:=\:0$.

\item {\em Choice of initial conditions}

As far as the choice of the initial conditions is concerned, the corrections
become smallest for large values of the initial occupation probability $p$
as can be seen in Fig. 5. The effect of correlated initial conditions
(an interpolating sequence of particles and holes) cannot be easily understood
and requires a more detailed analysis: for periodic boundary conditions, the
corrections become smaller for uncorrelated initial conditions (when the
initial concentration is kept the same) while for open boundary conditions,
the cluster configuration gives rise to smaller correction terms (cf. Tabs.
\ref{tab:icc} and \ref{tab:openex}).
So in general the most promising choice of initial conditions for the
determination of the scaling function and the scaling exponent
is the full lattice.

\item {\em Choice of boundary conditions}

The discussion which boundary conditions are more convenient for a numerical
examination of the scaling and the correction exponent
(the scaling functions are different for different boundary conditions)
is very difficult and depends on the other parameters as well.
If the initial occupation probability $p$ is smaller than $1$ and $\D'=0$
the corrections are
smaller for periodic boundaries.
For $\D'\not=0$ and $p=1$ however, the corrections are smaller for open
boundary conditions (cf. Tabs. \ref{tab:tsc} and \ref{tab:ftsc}). These cases
are interesting for the determination of the correction exponent.
Taking $p=1$ and
$\D'=0$ (the best choice of $\D'$ and $p$ for the determination of the
scaling exponent) the corrections are almost identical for different boundary
conditions. In Tabs. \ref{tab:tsc}
and \ref{tab:ftsc}
values for $z$ between $0.01$ and $0.1$ are given where the corrections are
exactly the same. For larger values of $z$, however, they become slightly
different (cf. paper II).
Taking into account that periodic boundaries usually allow a
Fourier transformation of the equations and
therefore a reduction of the number of degrees of freedom of the system,
it is more convenient to work with periodic
boundary conditions here.

\item {\em Value of the scaling variable $z$}

For the determination of the scaling exponent, one has to choose a
value of $z$ where the corrections are small. i.e. a value
around $z=0.6$ should be convenient. Since the zero of the correction function
occurs at a smaller value of $z$ for periodic boundary conditions, these
boundary
conditions better allow a Monte Carlo simulation: the smaller the
value of $z$ is, the smaller the simulated time can be made and the more
precise the results of the simulation are.
For the determination of the correction exponent the situation is quite
different. Here the correction function should be large in order to
allow a numerical fit of the simulated data. So one has to choose
small values of $z$ for a Monte Carlo simulation ($z\: \leq \:0.02$).
The BST-algorithm, however, converges better for large values of $z$
($z\: \geq \:0.2$).

\end{enumerate}

Whether these observations hold for more complicated chemical systems
still has to be investigated.
\end{itemize}

Summing up, we can say that the numerical treatment presented here completes
the picture of the finite-size scaling behaviour of the coagulation model.
Both, analytical and numerical methods can be succesfully applied to chemical
models. The results obtained in our work now open the way towards the
investigation of more complex chemical systems. The methods tested here
provide a powerful tool to gain deeper insight into non-equilibrium
physics.

\vspace{1cm}
\noindent {\large \bf Acknowledgements}
\\*[5mm]
\indent We would like to acknowledge Prof. V. Rittenberg
for helpful discussions.
One of us (H. S.) would like to thank the Deutscher Akademischer
Austauschdienst (DAAD) for financial support.
%

%
\pagebreak
\listoftables
\newcounter{fig_count}
\vspace{1cm}
\noindent {\Large \bf List of Figures}
\begin{list}
      {Fig. \arabic{fig_count}:}
      {\usecounter{fig_count}  \setlength{\leftmargin}{1.5cm}
               \setlength{\labelsep}{2mm}
               \setlength{\labelwidth}{1.3cm}}
\item First order correction function $F_{1}(z,\D')$
for periodic boundary conditions and different values of $\D'$.
\item Second order correction function $F_{2}(z,\D')$
for periodic boundary conditions and different values of $\D'$.
\item First order correction function $F_{1}(z,\D')$ for open boundary
conditions and different values of $\D'$.
\item Second order correction function $F_{2}(z,\D')$ for open boundary
conditions and different values of $\D'$.
\item Correction function $F_{2}(z,\D')$ for
periodic boundary conditions, $\D'=0$ and different values of the initial
occupation probability $p$.
\end{list}
%
\begin{table}[p]
\centering
\begin{tabular}{|r|r|r|c|} \hline
\multicolumn{1}{|c|}{$\D t$} &
\multicolumn{1}{c|}{$L$} &
\multicolumn{1}{c|}{$t_{max}$} &
\multicolumn{1}{c|}{Maximum error in \%.}\\
\hline
$5\cdot 10^{-3}$&$6$&$0.9$&$0.36$\\
$5\cdot 10^{-2}$&$200$&$10^{3}$&$0.42$\\
$5\cdot 10^{-1}$&$1000$&$25\cdot10^{3}$&$0.44$\\
\hline
\end{tabular}
\caption
[Maximum error of the simulated concentration for periodic boundaries
and $\D'=0$]
{Maximum error of the simulated concentration for periodic boundaries
and $\D'=0$}
\label{tab:taex}
\end{table}

\begin{table}[p]
\centering
\begin{tabular}{|r||r|r||r|r|} \hline
\multicolumn{1}{|c||}{} &
\multicolumn{2}{c||}{Scaling function $F_0(z)$} &
\multicolumn{2}{c|}{Corrections $F_2(z,0)$} \\
\hline
\multicolumn{1}{|c||}{z} &
\multicolumn{1}{c|}{anal.} &
\multicolumn{1}{c||}{M.C.} &
\multicolumn{1}{c|}{anal.} &
\multicolumn{1}{c|}{M.C.} \\
\hline
$0.01 $&$ 7.979$&$ 7.839\pm 0.005$&$-99.8$&$-103.2\pm0.3$\\
$0.02 $&$ 5.642$&$ 5.680\pm 0.004$&$-35.3$&$-45.0\pm0.3$\\
$0.03 $&$ 4.607$&$ 4.635\pm 0.004$&$-19.2$&$-24.3\pm0.3$\\
$0.04 $&$ 3.989$&$ 4.006\pm 0.003$&$-12.5$&$-15.0\pm0.3$\\
$0.05 $&$ 3.568$&$ 3.577\pm 0.003$&$-8.9$&$-10.1\pm0.3$\\
$0.06 $&$ 3.257$&$ 3.264\pm 0.003$&$-6.8$&$-7.4\pm0.2$\\
$0.07 $&$ 3.016$&$ 3.022\pm 0.003$&$-5.4$&$-5.8\pm0.2$\\
$0.08 $&$ 2.821$&$ 2.825\pm 0.003$&$-4.4$&$-4.7\pm0.2$\\
$0.09 $&$ 2.660$&$ 2.661\pm 0.003$&$-3.7$&$-3.7\pm0.2$\\
$0.10 $&$ 2.523$&$ 2.525\pm 0.003$&$-3.2$&$-3.2\pm0.2$\\
\hline
\end{tabular}
\caption
[Monte Carlo data and exact results for periodic boundaries and $\D'=0$]
{Monte Carlo data and exact results for periodic boundaries and $\D'=0$}
\label{tab:test}
\end{table}

\begin{table}[p]
\centering
\begin{tabular}{|r||r|r|r|r|r|c|}
\hline
\multicolumn{1}{|c||}{z} &
\multicolumn{1}{c|}{$\D'=0.75$} &
\multicolumn{1}{c|}{$\D'=0.5$} &
\multicolumn{1}{c|}{$\D'=0$} &
\multicolumn{1}{c|}{$\D'=-1$} &
\multicolumn{1}{c|}{$\D'=-2$} \\ \hline
$0.01 $&$ 7.95\pm 0.06 $&$ 7.87\pm 0.03 $&$7.98$&$ 7.97\pm 0.02 $&
$ 7.88\pm 0.03$\\
$0.02 $&$ 5.43\pm 0.04 $&$ 5.58\pm 0.02 $&$5.64$&$ 5.62\pm 0.02 $&
$ 5.57\pm 0.02$\\
$0.03 $&$ 4.50\pm 0.04 $&$ 4.55\pm 0.02 $&$4.61$&$ 4.58\pm 0.02 $&
$ 4.60\pm 0.02$\\
$0.04 $&$ 3.91\pm 0.04 $&$ 3.94\pm 0.02 $&$3.99$&$ 3.95\pm 0.02 $&
$ 3.96\pm 0.02$\\
$0.05 $&$ 3.50\pm 0.03 $&$ 3.55\pm 0.02 $&$3.57$&$ 3.54\pm 0.02 $&
$ 3.56\pm 0.02$\\
$0.06 $&$ 3.20\pm 0.03$&$ 3.25\pm 0.02 $&$3.26$&$ 3.23\pm 0.02 $&
$ 3.24\pm 0.02$\\
$0.07 $&$ 2.97\pm 0.03 $&$ 3.00\pm 0.02 $&$3.02$&$ 2.98\pm 0.02 $&
$ 3.00\pm 0.02$\\
$0.08 $&$ 2.78\pm 0.03 $&$ 2.80\pm 0.02 $&$2.82$&$ 2.80\pm 0.02 $&
$ 2.79\pm 0.02$\\
$0.09 $&$ 2.64\pm 0.03 $&$ 2.63\pm 0.02 $&$2.66$&$ 2.63\pm 0.01 $&
$ 2.63\pm 0.02$\\
$0.10 $&$ 2.49\pm 0.03 $&$ 2.51\pm 0.02 $&$2.52$&$ 2.51\pm 0.01 $&
$ 2.50\pm 0.02$\\
\hline
\end{tabular}
\caption
[Scaling function $F_0(z)$ for periodic boundaries]
{Scaling function $F_0(z)$ for periodic boundaries}
\label{tab:ts}
\end{table}

\begin{table}[p]
\centering
\begin{tabular}{|r||r|r|c|r|r|} \hline
\multicolumn{1}{|c||}{z} &
\multicolumn{1}{c|}{$\D'=0.75$} &
\multicolumn{1}{c|}{$\D'=0.5$} &
\multicolumn{1}{c|}{$\D'=0$} &
\multicolumn{1}{c|}{$\D'=-1$} &
\multicolumn{1}{c|}{$\D'=-2$} \\ \hline
$0.01 $&$ 471 \pm 6 $&$ 152.0 \pm 3.2 $&$0$&$ -59.7 \pm 2.4 $&$ -70 \pm 2.0
$\\
$0.02 $&$ 251 \pm 5 $&$  75.6 \pm 2.6 $&$0$&$ -28.8 \pm 2.0 $&$ -35 \pm 1.8
$\\
$0.03 $&$ 160 \pm 4 $&$  51.2 \pm 2.3 $&$0$&$ -18.3 \pm 1.9 $&$ -26 \pm 1.6
$\\
$0.04 $&$ 118 \pm 4 $&$  38.2 \pm 2.1 $&$0$&$ -11.7 \pm 1.8 $&$ -18 \pm 1.5
$\\
$0.05 $&$  93 \pm 4 $&$  28.8 \pm 2.0 $&$0$&$  -9.9 \pm 1.7 $&$ -16 \pm 1.5
$\\
$0.06 $&$  76 \pm 3 $&$  22.6 \pm 1.8 $&$0$&$  -7.9 \pm 1.6 $&$ -13 \pm 1.4
$\\
$0.07 $&$  65 \pm 3 $&$  20.0 \pm 1.7 $&$0$&$  -5.7 \pm 1.5 $&$ -11 \pm 1.4
$\\
$0.08 $&$  57 \pm 3 $&$  19.0 \pm 1.7 $&$0$&$  -5.4 \pm 1.5 $&$  -8 \pm 1.3
$\\
$0.09 $&$  49 \pm 3 $&$  18.0 \pm 1.6 $&$0$&$  -4.2 \pm 1.5 $&$  -7 \pm 1.3
$\\
$0.10 $&$  45 \pm 3 $&$  14.3 \pm 1.6 $&$0$&$  -4.8 \pm 1.4 $&$  -6 \pm 1.3
$\\
\hline
\end{tabular}
\caption
[First order correction function $F_1(z,\D')$ for periodic boundaries]
{First order correction function $F_1(z,\D')$ for periodic boundaries}
\label{tab:tfc}
\end{table}

\begin{table}[p]
\centering
\begin{tabular}{|r||r|r|r|r|r|} \hline
\multicolumn{1}{|c||}{z} &
\multicolumn{1}{c|}{$\D'=0.75$} &
\multicolumn{1}{c|}{$\D'=0.5$} &
\multicolumn{1}{c|}{$\D'=0$} &
\multicolumn{1}{c|}{$\D'=-1$} &
\multicolumn{1}{c|}{$\D'=-2$} \\ \hline
$0.01 $&$ -8212 \pm 156 $&$ -2137 \pm 74 $&$ -99.8  $&$ 448 \pm 53 $&$ 418
\pm 38 $\\
$0.02 $&$ -3197 \pm 124 $&$  -766 \pm 60 $&$ -35.3 $&$ 147 \pm 46 $&$ 148
\pm 33 $\\
$0.03 $&$ -1516 \pm 108 $&$  -438 \pm 52 $&$ -19.2 $&$ 56 \pm 42 $&$ 124
\pm 30 $\\
$0.04 $&$ -922 \pm 97 $&$  -289 \pm 48 $&$ -12.5 $&$ -16 \pm 39 $&$  61
\pm 28 $\\
$0.05 $&$  -634 \pm 90 $&$  -168 \pm 45 $&$ -8.9 $&$ -9 \pm 37 $&$  72
\pm 27 $\\
$0.06 $&$  -434 \pm 84 $&$   -97 \pm 42 $&$ -6.8 $&$ -20 \pm 36 $&$  50
\pm 26 $\\
$0.07 $&$  -332 \pm 80 $&$   -84 \pm 40 $&$ -5.4 $&$ -43 \pm 34 $&$  29
\pm 25 $\\
$0.08 $&$  -273 \pm 76 $&$   -97 \pm 39 $&$ -4.4 $&$ -35 \pm 33 $&$   4
\pm 24 $\\
$0.09 $&$  -177 \pm 73 $&$  -121 \pm 38 $&$ -3.7 $&$ -46 \pm 33 $&$  -4
\pm 23 $\\
$0.10 $&$  -183 \pm 71 $&$   -68 \pm 36 $&$ -3.2 $&$ -15 \pm 32 $&$  -7
\pm 23 $\\
\hline
\end{tabular}
\caption
[Second order correction function $F_2(z,\D')$ for periodic boundaries]
{Second order correction function $F_2(z,\D')$ for periodic boundaries}
\label{tab:tsc}
\end{table}

\begin{table}[p]
\centering
\begin{tabular}{|r||r|c|c|}
\hline
\multicolumn{1}{|c||}{z} &
\multicolumn{1}{c|}{$\D'=0.5$} &
\multicolumn{1}{c|}{$\D'=0$} &
\multicolumn{1}{c|}{$\D'=-1$} \\ \hline
$0.01$&$8.33\pm 0.04$&$8.34$&$8.27\pm 0.05$\\
$0.02$&$5.98\pm 0.03$&$6.01$&$6.00\pm 0.04$\\
$0.03$&$4.92\pm 0.03$&$4.97$&$4.99\pm 0.04$\\
$0.04$&$4.28\pm 0.02$&$4.35$&$4.41\pm 0.04$\\
$0.05$&$3.89\pm 0.02$&$3.93$&$3.96\pm 0.03$\\
$0.06$&$3.60\pm 0.02$&$3.62$&$3.65\pm 0.03$\\
$0.07$&$3.36\pm 0.02$&$3.38$&$3.40\pm 0.03$\\
$0.08$&$3.15\pm 0.02$&$3.18$&$3.18\pm 0.03$\\
$0.09$&$3.01\pm 0.02$&$3.02$&$3.01\pm 0.03$\\
$0.10$&$2.88\pm 0.02$&$2.89$&$2.85\pm 0.03$\\
\hline
\end{tabular}
\caption
[Scaling function $F_0(z)$ for open boundaries]
{Scaling function $F_0(z)$ for open boundaries}
\label{tab:fts}
\end{table}

\begin{table}[p]
\centering
\begin{tabular}{|r||r|c|r|} \hline
\multicolumn{1}{|c||}{z} &
\multicolumn{1}{c|}{$\D'=0.5$} &
\multicolumn{1}{c|}{$\D'=0$} &
\multicolumn{1}{c|}{$\D'=-1$} \\ \hline
$0.01$&$132\pm 3$&$0$&$-58\pm 4$\\
$0.02$&$ 65\pm 2$&$0$&$-33\pm 4$\\
$0.03$&$ 44\pm 2$&$0$&$-25\pm 3$\\
$0.04$&$ 36\pm 2$&$0$&$-24\pm 3$\\
$0.05$&$ 27\pm 2$&$0$&$-17\pm 3$\\
$0.06$&$ 21\pm 2$&$0$&$-15\pm 3$\\
$0.07$&$ 18\pm 2$&$0$&$-13\pm 3$\\
$0.08$&$ 16\pm 2$&$0$&$ -9\pm 3$\\
$0.09$&$ 13\pm 2$&$0$&$ -7\pm 3$\\
$0.10$&$ 11\pm 2$&$0$&$ -4\pm 3$\\
\hline
\end{tabular}
\caption
[First order correction function $F_1(z,\D')$ for open boundaries]
{First order correction function $F_1(z,\D')$ for open boundaries}
\label{tab:ftfc}
\end{table}

\begin{table}[p]
\centering
\begin{tabular}{|r||r|r|r|} \hline
\multicolumn{1}{|c||}{z} &
\multicolumn{1}{c|}{$\D'=0.5$} &
\multicolumn{1}{c|}{$\D'=0$} &
\multicolumn{1}{c|}{$\D'=-1$} \\ \hline
$0.01$&$-1845\pm 56$&$-99.7$&$355\pm 97$\\
$0.02$&$ -648\pm 45$&$-35.3$&$180\pm 83$\\
$0.03$&$ -386\pm 40$&$-19.2$&$173\pm 76$\\
$0.04$&$ -308\pm 36$&$-12.5$&$215\pm 71$\\
$0.05$&$ -207\pm 34$&$ -8.9$&$123\pm 68$\\
$0.06$&$ -144\pm 32$&$ -6.8$&$125\pm 66$\\
$0.07$&$ -112\pm 31$&$ -5.4$&$ 97\pm 63$\\
$0.08$&$ -113\pm 30$&$ -4.4$&$ 29\pm 62$\\
$0.09$&$  -71\pm 29$&$ -3.7$&$  7\pm 60$\\
$0.10$&$  -55\pm 28$&$ -3.2$&$-39\pm 59$\\
\hline
\end{tabular}
\caption
[Second order correction function $F_2(z,\D')$ for open boundaries]
{Second order correction function $F_2(z,\D')$ for open boundaries}
\label{tab:ftsc}
\end{table}

\begin{table}[p]
\centering
\begin{tabular}{|r||r|} \hline
\multicolumn{1}{|c||}{$\D'$} &
\multicolumn{1}{c|}{$a(\D',1)$} \\ \hline
$-2.00 $&$ -0.512\pm 0.060$\\
$-1.00 $&$ -0.414\pm 0.083$\\
$ 0.50 $&$  0.894\pm 0.070$\\
$ 0.75 $&$  2.831\pm 0.130$\\
\hline
\end{tabular}
\caption
[First correction term $a(\D',1)$ of the large-time expansion for
periodic boundaries]
{First correction term $a(\D',1)$ of the large-time expansion for
periodic boundaries}
\label{tab:ltime}
\end{table}

\begin{table}[p]
\centering
\begin{tabular}{|r||r|r||r|r|} \hline
\multicolumn{1}{|c||}{} &
\multicolumn{2}{c||}{Scaling function $F_0(z)$} &
\multicolumn{2}{c|}{Corrections $F_2(z,0)$} \\
\hline
\multicolumn{1}{|c||}{} &
\multicolumn{1}{c|}{$p=0.5$} &
\multicolumn{1}{c||}{$AA\nil \nil \ldots$} &
\multicolumn{1}{c|}{$p=0.5$} &
\multicolumn{1}{c|}{$AA\nil \nil \ldots$} \\
\multicolumn{1}{|c||}{z} &
\multicolumn{1}{c|}{anal.} &
\multicolumn{1}{c||}{M.C.} &
\multicolumn{1}{c|}{anal.} &
\multicolumn{1}{c|}{M.C.} \\
\hline
$0.01 $&$7.979$&$7.967\pm 0.008$&$ -212.5$&$ -836.7\pm 12.6$\\
$0.02 $&$5.642$&$5.623\pm 0.007$&$ -106.3$&$ -304.8\pm 10.7$\\
$0.03 $&$4.607$&$4.607\pm 0.006$&$ -70.8$&$ -164.4\pm 9.7$\\
$0.04 $&$3.989$&$3.988\pm 0.006$&$ -53.1$&$ -109.8\pm 9.0$\\
$0.05 $&$3.570$&$3.556\pm 0.006$&$ -42.5$&$ -77.2\pm 8.6$\\
$0.06 $&$3.257$&$3.245\pm 0.005$&$ -35.4$&$ -61.7\pm 8.2$\\
$0.07 $&$3.016$&$3.006\pm 0.005$&$ -30.4$&$ -48.8\pm 7.9$\\
$0.08 $&$2.821$&$2.808\pm 0.005$&$ -26.6$&$ -40.6\pm 7.6$\\
$0.09 $&$2.660$&$2.647\pm 0.005$&$ -23.6$&$ -27.7\pm 7.4$\\
$0.10 $&$2.523$&$2.517\pm 0.005$&$ -21.3$&$ -17.6\pm 7.2$\\
\hline
\end{tabular}
\caption
[Influence of the initial conditions on the scaling properties for
periodic boundaries and $\D'=0$]
{Influence of the initial conditions on the scaling properties for
periodic boundaries and $\D'=0$}
\label{tab:icc}
\end{table}
\begin{table}[p]
\centering
\begin{tabular}{|r||l|l|l||r|l|l|} \hline
\multicolumn{1}{|c||}{} &
\multicolumn{1}{c|}{$1+x$} &
\multicolumn{2}{c||}{Scaling function $F_{0}(z)$} &
\multicolumn{1}{c|}{$-2-y$} &
\multicolumn{2}{c|}{Corrections $F_{2}(z)$}\\
\hline
\multicolumn{1}{|c||}{z} &
\multicolumn{1}{c|}{extr.} &
\multicolumn{1}{c|}{extr.} &
\multicolumn{1}{c||}{analytical}&
\multicolumn{1}{c|} {extr.}&
\multicolumn{1}{c|}{extr.} &
\multicolumn{1}{c|}{analytical}  \\
\hline
0.1 & 0.004547  & 2.5210     & 2.5231          & -0.134 & -2.776 & -3.154 \\
0.2 & 0.002986  & 1.784283   & 1.784286        & -0.011 & -1.1011 & -1.1017 \\
0.3 & 0.0000353 & 1.46043902 & 1.46043901      & -0.005 & -0.51697 & -0.51697
\\
0.4 & 0.0000003 & 1.27857491 & 1.27856700      & 0.005  & -0.22442 & -0.22246
\\
0.5 & 0.0000477 & 1.169713390 & 1.169713392   & 0.026   & -0.07196 & -0.07196
\\
0.6 & -0.000002 & 1.103560880 & 1.103560906   & 0.145   & -0.00291 & -0.00291
\\
0.7 & 0.0000304 & 1.0632169524 & 1.0632169522 & -0.003  & 0.023694 & 0.023694
\\
0.8 & 0.0000726 & 1.0385928840 & 1.0385928831 & 0.005   & 0.030097 & 0.030097
\\
0.9 & 0.0000522 & 1.02356074810 & 1.02356074816
& -0.002 & 0.0279317 & 0.0279317 \\
1.0 & 0.0000988 & 0.01438377205 & 0.01438377206
& 0.006 & 0.0228893 & 0.0228893 \\
\hline
\end{tabular}
\caption{Extrapolants from finite lattices ($L=2,\dots,9$) for periodic
boundary conditions}
\label{tab:extra}
\end{table}
%
\begin{table}[p]
\centering
\begin{tabular}{|c||r|r||r|r|} \hline
\multicolumn{1}{|c||}{} &
\multicolumn{2}{c||}{Scaling function $F_0(z)$} &
\multicolumn{2}{c|}{Corrections $F_2(z)$} \\
\hline
\multicolumn{1}{|c||}{z} &
\multicolumn{1}{c|}{$p=0.5$} &
\multicolumn{1}{c||}{$AA\nil\nil\ldots$} &
\multicolumn{1}{c|}{$p=0.5$} &
\multicolumn{1}{c|}{$AA\nil\nil\ldots$} \\
\multicolumn{1}{|c||}{} &
\multicolumn{1}{c|}{anal.} &
\multicolumn{1}{c||}{extrapol.} &
\multicolumn{1}{c|}{anal.} &
\multicolumn{1}{c|}{extrapol.} \\
\hline
0.01 & 8.342226 & 8.344116 & -1950.15 & -1050.17 \\
0.02 & 6.005276 & 6.005276 & -726.78  & -381.02 \\
0.04 & 4.352803 & 4.352803 & -275.60  & -144.03 \\
0.07 & 3.379100 & 3.379100 & -127.93  & -66.66 \\
0.10 & 2.886513 & 2.886513 & -79.08   & -41.12 \\
0.20 & 2.147403 & 2.147403 & -31.71   & -16.41 \\
0.30 & 1.818125 & 1.818125 & -18.94   & -9.79 \\
0.40 & 1.617195 & 1.617195 & -13.21   & -6.81 \\
0.50 & 1.475872 & 1.475872 & -9.83    & -5.04 \\
0.60 & 1.369969 & 1.369969 & -7.50    & -3.82 \\
0.70 & 1.288534 & 1.288534 & -5.77    & -2.91 \\
0.80 & 1.225287 & 1.225287 & -4.45    & -2.23 \\
0.90 & 1.175981 & 1.175981 & -3.44    & -1.70 \\
1.00 & 1.137489 & 1.137489 & -2.66    & -1.30 \\
\hline
\end{tabular}
\caption{Influence of the initial conditions on the scaling properties
for open boundaries and $\D'=0$}
\label{tab:openex}
\end{table}

\begin{thebibliography}{99}
\bibitem{Pap1}     K. Krebs, M. Pfannm\"uller and B. Wehefritz,
                   {\sl Finite-Size Scaling Studies of Reaction-Diffusion
                        Systems, Part I: Periodic Boundary Conditions},
                   {\em University of Bonn preprint} BONN HE-93-51 (1993).
\bibitem{Pap2}     H. Hinrichsen, K. Krebs, M. Pfannm\"uller and B. Wehefritz,
                   {\sl Finite-Size Scaling Studies of Reaction-Diffusion
                        Systems, Part II: Open Boundary Conditions},
                   {\em University of Bonn preprint} BONN HE-94-01 (1994).
\bibitem{Alcaraz}  F.C. Alcaraz, M. Droz, M. Henkel and V. Rittenberg,
                   {\sl Reaction-Diffusion Processes, Critical Dynamics and
                   Quantum Chains},
                   {\em University of Gen\`eve preprint UGVA-DPT} 1992/12-799,
                    to appear in {\em Ann. Phys.} (1994).
\bibitem{Henkel_unp} F. C. Alcaraz, M. Henkel and V. Rittenberg
                    {\em unpublished}.
\bibitem{Argyrakis} P. Argyrakis,
                   {\em Computers in Physics} {\bf 6}:525 (1992).
\bibitem{ben-Avraham} D. ben-Avraham,
                   {\em J. Chem. Phys.} {\bf 88}:941 (1988).
\bibitem{Doering}  C.R. Doering and D. ben-Avraham,
                   {\em Phys. Rev. A} {\bf 38}:3035 (1988).
\bibitem{Bulirsch} R. Bulirsch and J. Stoer,
                    {\em Numer. Math.} {\bf 6}:413 (1964).
\bibitem{Malte}   P. Christe and M. Henkel,
                   {\sl Introduction to Conformal Invariance and
                   its Application to Critical Phenomena},
                   (Springer, Berlin, Heidelberg, New York, 1993), chapter 3.
\end{thebibliography}
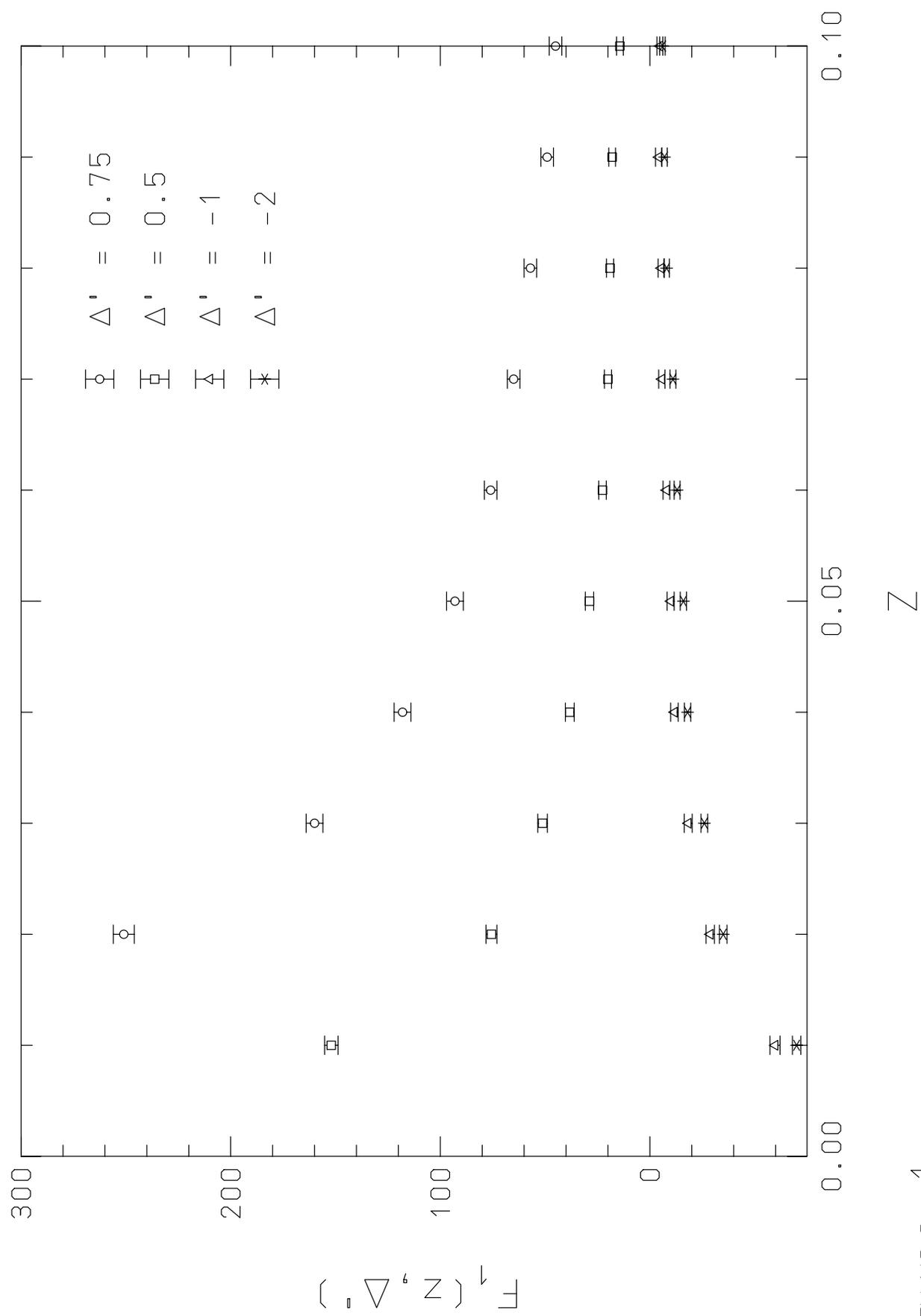
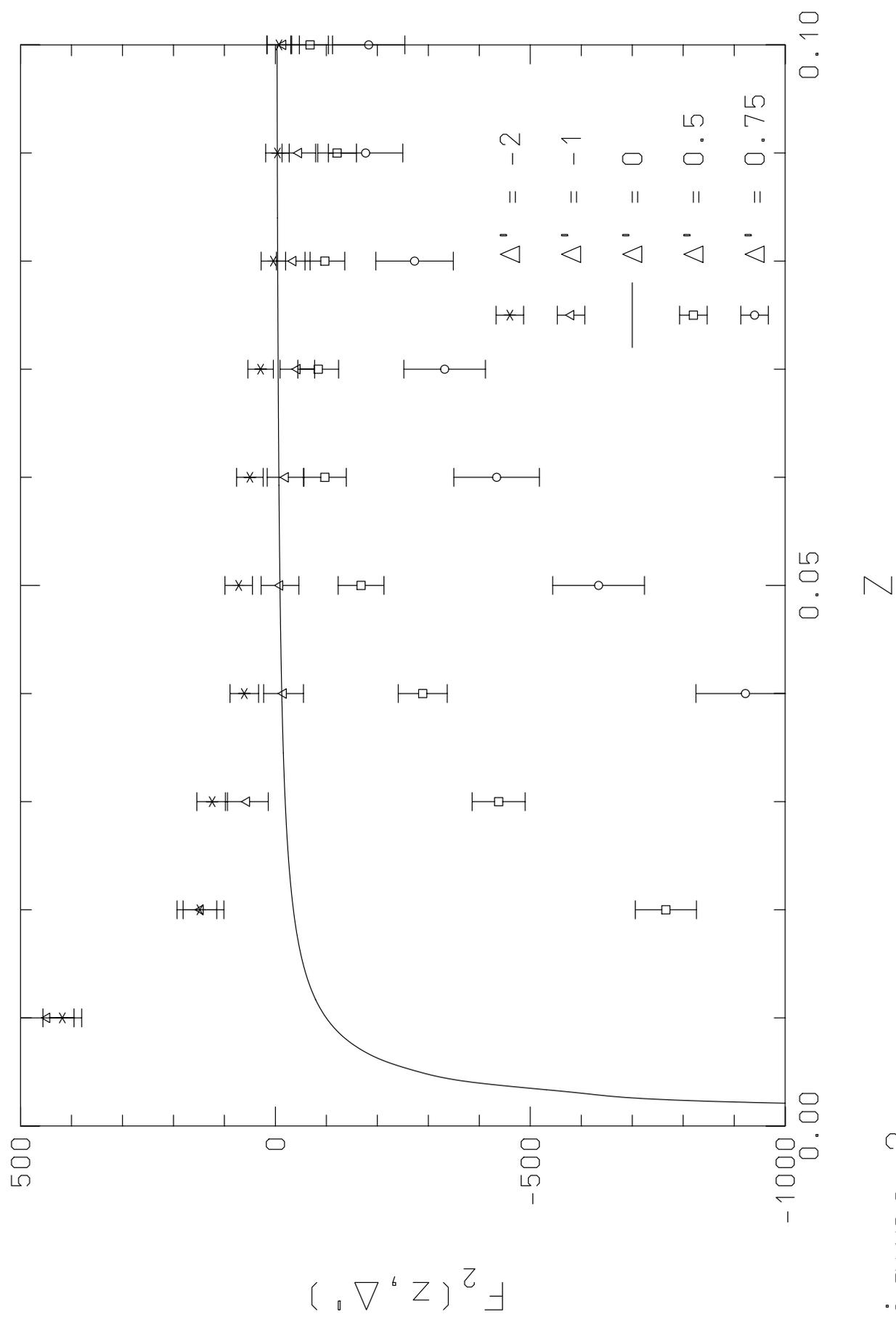
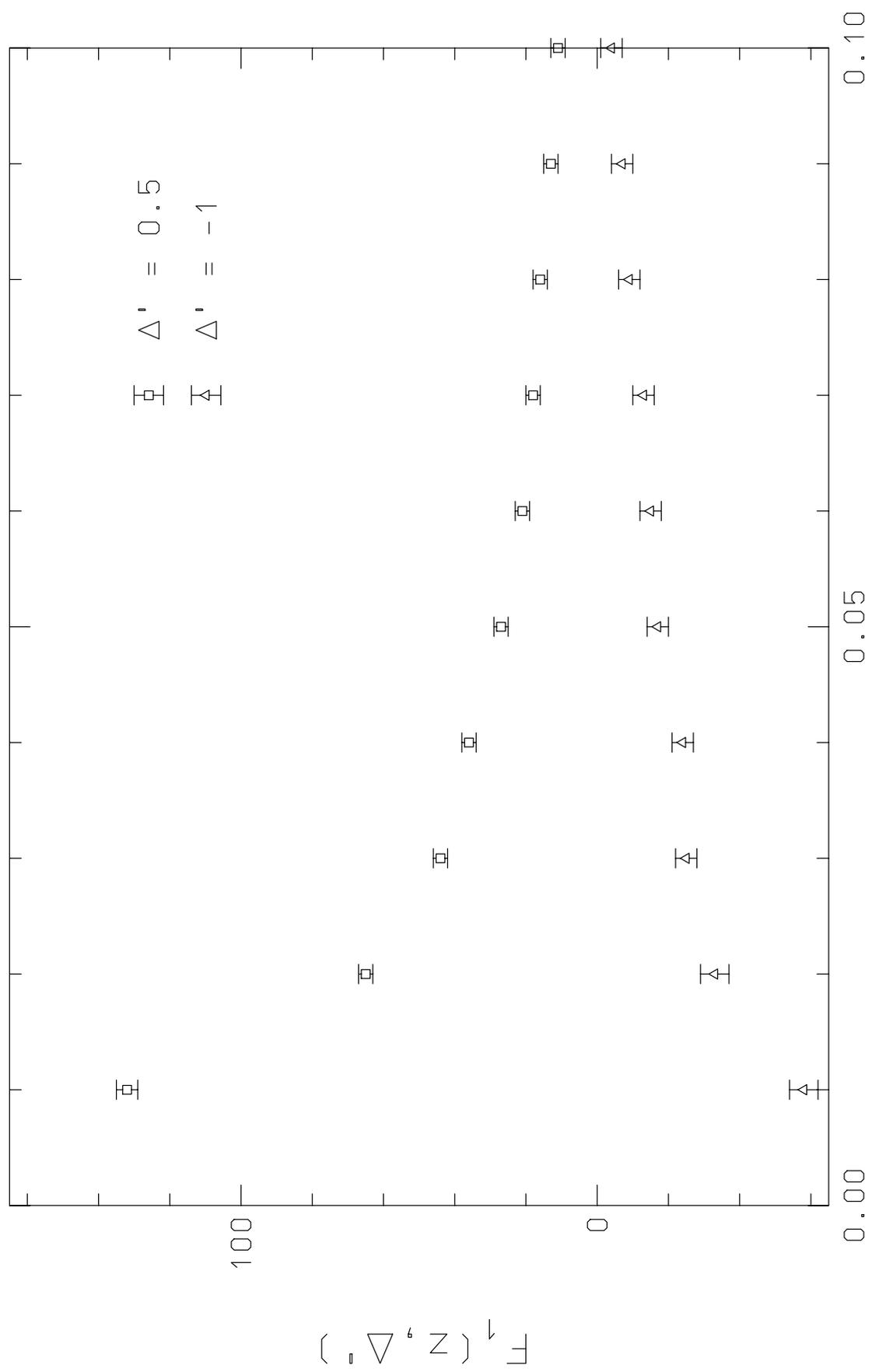
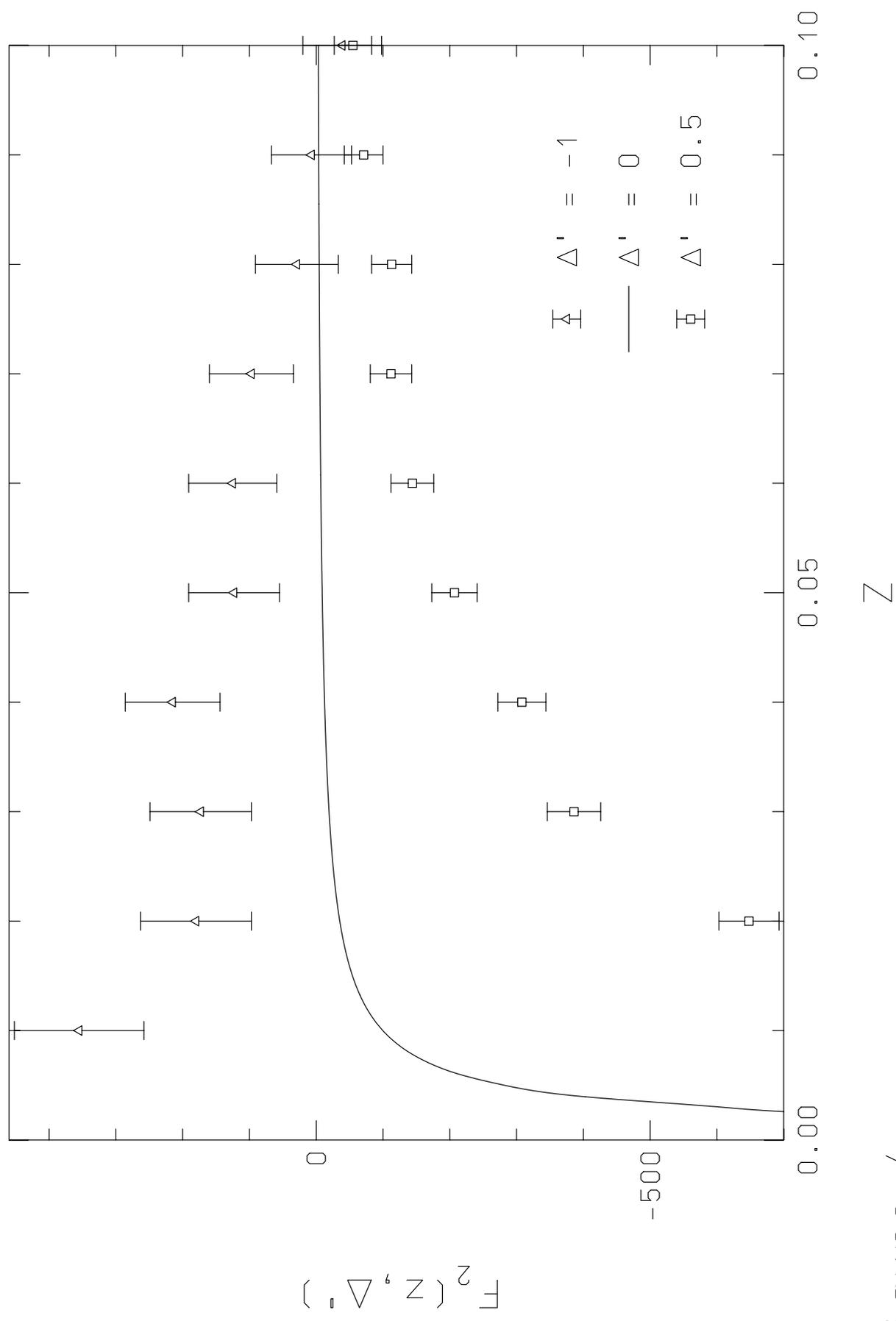
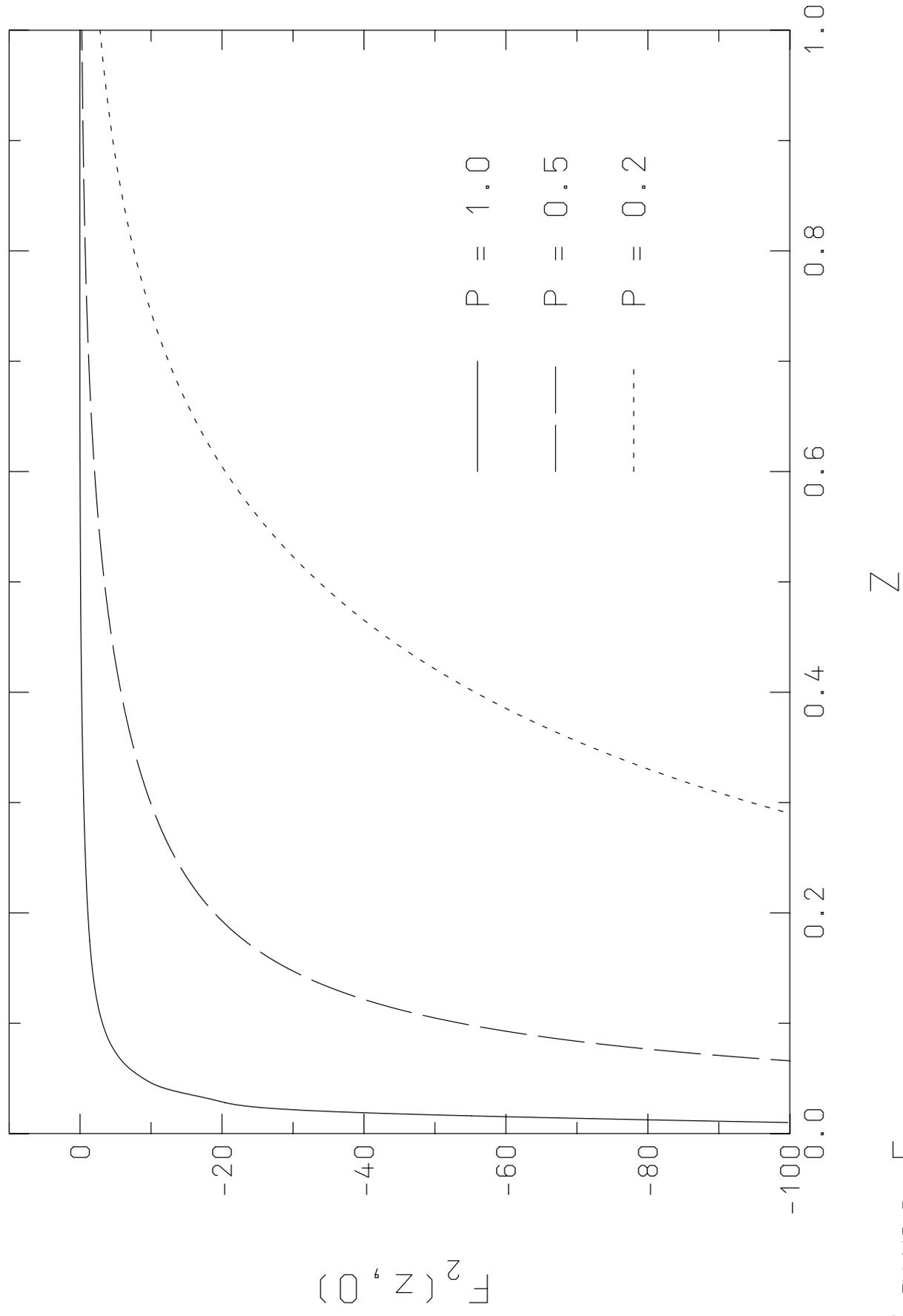
\end{document}